\documentstyle[aps,12pt,preprint]{revtex}
\begin{document}
\title{\bf The Role of Spin Anisotropy in the Unbinding of Interfaces}
\author{C. Micheletti and J. M. Yeomans}
\address{Department of Physics, Theoretical Physics, University of Oxford,\\1
Keble Road, Oxford OX1 3NP, U.K.}
\maketitle
\begin{abstract}
We study the ground state of a classical X-Y model with $p \ge 3$-fold
spin anisotropy $D$ in a uniform external field, $H$. An interface is
introduced into the system by a suitable choice of boundary
conditions. For large $D$, as $H \to 0$, we prove using an expansion
in $D^{-1}$ that the interface unbinds from the surface through an
infinite series of layering transitions. Numerical work shows that the
transitions end in a sequence of critical end points.
\vspace{1. truecm}
\bgroup\draft\pacs{PACS numbers: 0550; 6470; 7540D}\egroup
\end{abstract}
When an interface unbinds from a surface it can do so through a first
or second order transition or, on a lattice, via a sequence of first
order layering transitions. In general the nature of the unbinding, or
wetting, transition depends on the details of the microscopic
interactions and external parameters such as the temperature and
applied magnetic field~\cite{oliveira,pandit.et.al}.

In a discrete spin model at zero temperature long-range interactions
are needed to stabilise series of first order layering transitions.
However, if the interactions are short range, an infinite sequence of
such transitions can occur at finite temperatures~\cite{Duxbury.and.J}. This
results from competition between the localising effect of the binding
potential and the entropic terms in the free energy which favour a
delocalised interface.

In this paper we aim to discuss the role of a hithertofore unexplored
parameter on the unbinding transition: the spin anisotropy. We shall
show that, as discrete spins soften, layering transitions can be
stabilised in simple, short-range clock models, even at zero
temperature. An expansion in inverse spin anisotropy allows us to
prove that an infinite sequence of layering phase transitions exist.
Moreover, because the interesting features occur at zero temperature it
is possible to follow the phase diagram numerically for all values
of the spin anisotropy. In particular we are able to demonstrate how
the boundaries between the different interface phases end in
critical end points and to pinpoint these with considerable precision.

We consider the classical X-Y model in a magnetic field $H$ with
$p$-fold anisotropy $D$. The model is defined by the Hamiltonian
\begin{equation}
\label{eqn:hamiltonian}
{\cal H} = \sum_{i=1}^N \left\{ -J \cos  (\theta_{i-1} - \theta_i) - H (
\cos \theta_i -1) - D  (\cos p\theta_i -1)/p^2 \right\}
\end{equation}
where $i$ labels the spins on a one-dimensional lattice and
$\theta_i$ can take values between 0 and $2\pi$. An interface is
imposed on the system by choosing $\theta_0= 2 \pi / p$, and
$\theta_N=0$ and letting $N \to \infty$. If the interface lies
to the right of the $n^{\rm th}$ spin from the end of the chain (that is,
for $\theta_0$ fixed, between $i=n-1$ and $i=n$) the corresponding
interface phase will be labelled $\langle n \rangle$.

For infinite $D$ the Hamiltonian~(\ref{eqn:hamiltonian})
 describes a $p$-state clock model.
For $H >0$ the interface is bound to the surface in state $\langle 1
\rangle$.
The point $(H~=~0~,~D=\infty)$ is a
multiphase point where the interface has the same energy whatever its
position on the lattice.
As the anisotropy is reduced from infinity it becomes, for $p \ge 3$,
energetically
favourable for the spins to relax from their clock positions.
This results from competition
between the field term, which favours an interface in position
$\langle 1 \rangle $ and the exchange interaction which prefers to
minimise the angle between the spins and hence favours a free
interface. The result of this is that the degeneracy at the multiphase
point is broken and the interface unbinds from the surface through an
infinite series of first order layering transitions $\langle 1
\rangle, \langle 2 \rangle, \langle 3 \rangle \dots$ .

For large $D$ the existence of these transitions can be proven using
an expansion in $D^{-1}$~\cite{F.and.J}.
Writing
\begin{equation}
\label{eqn:angexp}
 \theta_i= \theta_i^0+ \tilde{\theta_i}
\end{equation}
where $\theta_i^0 \equiv \theta_i(D=\infty)$,
and keeping only terms quadratic in the angles and their differences
the Hamiltonian~(\ref{eqn:hamiltonian}) becomes
\begin{eqnarray}
\label{eqn:hamexp}
{\cal H} = \sum_{i=1}^\infty& &  \left\{  -  J c_i - H \cos(\theta_i^0-1)  +  J
c_i
\{ \tilde{\theta}_{i-1} - \tilde{\theta_i} + {s_i/c_i}\}^2/2 - J
s_i^2/2 c_i+\right. \nonumber \\
& & H \cos \theta_i^0\{\tilde{\theta_i}  + \tan \theta_i^0\}^2/2
\left. - H
\sin^2 \theta_i^0/ 2 \cos \theta_i^0+ D \tilde{\theta_i}^2/2 \right\}
\end{eqnarray}
where
\begin{equation}
\label{eqn:definitions}
 s_i=\sin(\theta_{i-1}^0-\theta_i^0), \quad c_i=
\cos(\theta_{i-1}^0-\theta_i^0) .
\end{equation}

The equilibrium values of the $\tilde{\theta_i}$ are given by
minimising the Hamiltonian (\ref{eqn:hamexp}). This leads to linear
recursion equations
\begin{equation}
\label{eqn:recursion}
 -J c_i \{ \tilde{\theta}_{i-1} - \tilde{\theta}_i + {s_i/c_i}\} + J
c_{i+1} \{\tilde{\theta}_i - \tilde{\theta}_{i+1} + {s_{i+1}/c_{i+1}}\} +
H \cos \theta_i^0 ( \tilde{\theta}_i+ \tan \theta_i^0) + D
\tilde{\theta_i} =0.
\end{equation}
If the full Hamiltonian~(\ref{eqn:hamiltonian}) is used
non-linearities appear in the recursion
relations~(\ref{eqn:recursion}).
 However, these do not affect the leading order
terms needed for the subsequent calculations.

Leading order corrections to the values of the spins can easily be
read off from the recursion equations~(\ref{eqn:recursion}). For the
phase $\langle n \rangle$
\begin{eqnarray}
\label{eqn:angles}
& & \vdots \nonumber \\
& & \tilde{\theta}_{n+2}  =  ({J / D})^3 \sin (2 \pi / p) + {\cal
O}({1 / D^4}) \nonumber \\
& & \tilde{\theta}_{n+1}  =  ({J / D})^2 \sin (2 \pi / p) + {\cal
O}({1 / D^3})\nonumber \\
& & \tilde{\theta}_{n}  =  (J / D) \sin (2 \pi / p) + {\cal
O}({1 / D^2})\nonumber \\ [0.5cm]
& & \tilde{\theta}_{n-1}  =  -{J / D} \sin (2 \pi / p) + {\cal
O}({H / D}, 1/D^2)\nonumber \\
& &\tilde{\theta}_{n-2}  =  -({J / D})^2 \sin (2 \pi / p) + {\cal
O}({H / D}, 1/D^3)\nonumber \\
& & \vdots \nonumber \\
& & \tilde{\theta}_{1}  =  -({J / D})^{n-1} \sin (2 \pi / p) + {\cal
O}({H / D}, 1/D^n)
\end{eqnarray}
The final result for the interface phase boundaries derived below will
demonstrate
that it is consistent to neglect tems ${\cal O}({H / D})$
in~(\ref{eqn:angles}). Note that for $p=2$, for $D$ large, the spins do
not lower their energy by canting.

Using the harmonic approximation~(\ref{eqn:hamexp}) the energy
differences $E_{\langle n \rangle} - E_{\langle
n-1 \rangle}$ between neighbouring interface states can be calculated.
 Let $\langle n-1
\rangle$ have spins $\{ \alpha_i\}$ with $\alpha_1$ the surface spin
and $\langle n \rangle$ have spins $\{ \beta_i\}$ with $\beta_0$ the
surface spin. Then in both cases the interface lies between $i=n-1$
and $i=n$ and $\alpha_i^0 \equiv \beta_i^0$. Using this labelling and
the recursion equations (\ref{eqn:recursion}) some algebra leads to
the result
\begin{eqnarray}
\label{eqn:endiff1}
 & &  E_{\langle n \rangle} - E_{\langle n-1 \rangle} = -H \left\{ \cos (2 \pi
/ p) -1\right\} + H \tilde{\beta}_1 \sin \beta_1^0/2 - J
\tilde{\beta}_1 \tilde{\alpha}_2/2 \quad \hskip 2.0cm  n \ge 3 \\
\label{eqn:endiff2}
 & &  E_{\langle 2 \rangle} - E_{\langle 1 \rangle} = -H \left\{ \cos (2 \pi
/ p) -1\right\} + H \tilde{\beta}_1 \sin \beta_1^0/2 - J \cos (2 \pi /
p) \tilde{\beta}_1 \tilde{\alpha}_2/2 + J \sin (2 \pi / p)
\tilde{\beta}_1/2
\end{eqnarray}
These formulae are exact for the quadratic
Hamiltonian~(\ref{eqn:hamexp}).
Higher order terms in the full Hamiltonian~(\ref{eqn:hamiltonian})
appear as higher order corrections.

Substituting in the values for the surface spins from
(\ref{eqn:angles}) gives, for $n \ge 2$,
\begin{equation}
\label{eqn:energydiff}
 E_{\langle n \rangle} - E_{\langle n-1 \rangle} = - H (\cos ( 2 \pi
/ p) -1) - { J^{2n -2}\sin^2 (2 \pi/ p) / (2 D^{2n-3})}  +
{\cal O}(1/D^{2n-2})
\end{equation}
It follows immediately from~(\ref{eqn:energydiff}) that the boundary
between phases $\langle n-1 \rangle$ and $\langle n \rangle$ is given
to leading order by

\begin{equation}
\label{eqn:boundary}
 H_{\langle n-1 \rangle : \langle n \rangle} = J^{2n -2} \sin^2 (2 \pi / p)
\left\{ 2  \left(1 - \cos (2 \pi
/ p) \right) D^{2n -3}\right\}^{-1}
\end{equation}
indicating that the unbinding proceeds via an infinite series of
phases $\langle n \rangle$ of widths ${\cal O}( {1 / D^{2n-3}})$.

These results were confirmed numerically for the case $p=3$ by studying
iteratively the
equations which minimize the energy~(\ref{eqn:hamiltonian}). The
numerical approach allowed us to obtain the interface phase diagram for all
values of $D$ which is shown in Figure 1. The first order boundaries
between  the different interface phases end at a series of critical end
points at
\begin{eqnarray*}
\langle 1 \rangle & : & \langle 2 \rangle  \quad D^*_{1,2}=1.1268 \pm
0.0003  \quad H^*_{1,2} = 0.2566 \pm 0.0009\\
\langle 2 \rangle & : & \langle 3 \rangle  \quad D^*_{2,3}=0.9360 \pm
0.0003  \quad H^*_{2,3} = 0.04357 \pm 0.00001\\
\langle 3 \rangle & : & \langle 4 \rangle  \quad D^*_{3,4}=0.7281 \pm
0.0005  \quad H^*_{3,4} = 0.01029 \pm 0.00004\\
\langle 4 \rangle & : & \langle 5 \rangle  \quad D^*_{4,5}=0.5931 \pm
0.0003  \quad H^*_{4,5} = 0.00295 \pm 0.00002\\
\vdots
\end{eqnarray*}
These were identified as the points where both the energy $E$ and
its partial derivative with respect to $H$ become the same in the two phases.
Assuming that $D^*_{n,n+1}$ and $H^*_{n,n+1}$ have a power law
dependence on $n$, $(D^*_\infty,H^*_\infty)=(0,0)$ is consistent
with the data. However, it was only possible to obtain results for $n
\le 5$, and so we cannot be confident that
this is the true asymptotic behaviour.

For $D=0$ the interface shape varies continuously
from being a domain wall at the surface for $H$ large to a uniform
spiral for $H=0$.
\vskip1.0cm
To summarise, we have shown that the softening of discrete spins
provides a mechanism, somewhat analogous to temperature, which can
stabilise interface layering transitions. An expansion in inverse spin
anisotropy was used to demonstrate that an infinite number of such transitions
exist at large $D$. Because the transitions take place at zero
temperature it was possible to obtain good numerical estimates of the
end points of the first order transition lines.

The model we describe is widely used in the theoretical description of
magnetism in the rare-earth metals and
compounds~\cite{rare.earths1,rare.earths2}.
The results
presented here may be of relevance to attempts to model mutilayers
comprised of these compounds~\cite{review}.
Moreover, we hope that the system will have
theoretical applicability in that it represents one of the simplest
models where expansion about a multiphase point is feasible.
Aubry~\cite{Aubry} has recently pointed out that such multiphase
points (which he calls anti-integrable limits) also exist in
electronic systems and arrays of non-linear coupled oscillators. It
may be possible to use extensions of the technique presented here to
treat these systems.
\vskip1.3cm
{\Large Aknowledgements}
\vskip 0.3cm
We are grateful to Dr. F. Seno for useful discussions. JMY aknowledges
support from an E.P.S.R.C. Advanced Fellowship and CM from an E.P.S.R.C.
studentship and from Fondazione ``A. Della Riccia'', Firenze.

\vskip 1.8cm
{\Large Figure Caption}
\vskip 0.5cm
Fig 1: Phase diagram of the classical X-Y model in a
magnetic field, $H$, with 3-fold spin anisotropy, $D$, and an imposed
interface. There are an infinite number of interface layering
transitions. The first order boundaries between them terminate in
critical end points.
\end{document}